# On unconventional fall of rainbow spring

Jie Liu[1,*], Xing Wang[1], Jiongzhao Liang[1], Bing Liu[1], Wei Wei[2],
Huimin Shi[1], Guilin Wen[1,*], Yi Min Xie[3]

[1]*Center for Research on Leading Technology of Special Equipment, School of Mechanical and Electric Engineering, Guangzhou University, Guangzhou 510006, P. R. China*

[2]*Department of Electronic and Information Engineering, The Hong Kong Polytechnic University, Hung Hom, Kowloon, Hong Kong, P. R. China*

[3]*Centre for Innovative Structures and Materials, School of Engineering, RMIT University, Melbourne 3001, Australia*

*Correspondence and requests for materials should be addressed to G.L.W. (email: glwen@gzhu.edu.cn) and J.L. (email: jliu@gzhu.edu.cn)

In this study we experimentally show that a stretched rainbow spring under gravity or extra weight may exhibit unconventional fall motion. Specially, when the rainbow spring is released from a high place, its lower end remains stationary until the spring wires stack together and then all the parts of the rainbow spring falls down together. We utilize a high-speed camera to record the fall process of one plastic and one metal rainbow spring under different loading conditions to systematically investigate this unconventional physical phenomenon. We use the time of the elastic wave propagating the length of the spring to predict the duration of the lower end of the rainbow spring remaining still. The findings from this study elucidate this physical phenomenon which has potential for the areas requiring temporary absolute space positioning.

Temporary absolute space positioning phenomenon can be commonly found in the movement of birds, i.e. their heads keep still in the air for a while in each step [1-3]. The capacity is mainly to improve the stability of gaze within the movement, thereby improving the accuracy of the predation, especially for catching moving targets. This natural phenomenon may be imitated to solve the high resolution satellite positioning problem [4,5].

In addition to the birds, there is another object whose movement may also have the temporary absolute space positioning phenomenon. In our daily life of the macro world, we often have such an impression that the whole object will fall together when it is released from a high place due to gravity. This may be because of our focus being only on the fall of the rigid body [6], therefore leading to the misunderstanding that all the physical objects fall down together due to gravity.

However, the same may not happen to a structure that is deformable inside itself, e.g. the rainbow spring [7]. During its movement, it may exhibit the interesting phenomenon aforementioned.

In this Letter we report for the first time the observation of the unconventional fall of stretched rainbow springs under gravity or other static loading using high-speed camera. We investigate two processes of the fall: a) the lower end of the spring holds still in the air for a while (means there is a temporary absolute space positioning) and its upper parts move down, and (b) the whole spring falls together. We reveal that the unconventional fall in the first process is due to the transformation from elastic potential energy, produced by gravity and/or other static loading, to kinetic energy. The time of the lower end holds still, $t$, is predicted by using the duration of the elastic wave propagating the length of the spring, $t_w$. We emphasize that the reported unconventional physical phenomenon may be observed in other objects as well.

First, we consider a helical rainbow spring, shown in Figure 1, twined from wire whose central axis constructs a left-handed helix. The helical spring is statically loaded by self-gravity extra weight and its length in static equilibrium is $L$. The wire diameter is negligibly small compared to $L$. The static axial tension at the end of the rainbow spring, $T$, consists of two parts, i.e. gravity and external loading, as

$$T = (m + M)g \tag{1}$$

where $m$ is the total mass of the spring, $g$ the gravity acceleration, and $M$ the mass of the weight at the bottom of the spring.

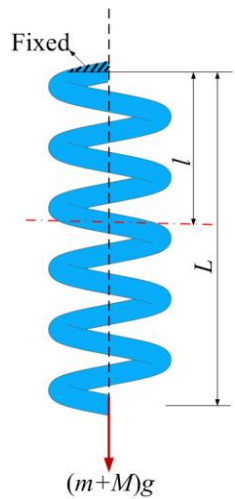

$(m+M)g$

**FIG. 1.** Statically loaded helical spring. The rainbow spring may be loaded by the gravity or other static loading at the bottom.

Let $T = m_0 g$, yields the equivalent mass, as

$$m_0 = m + \frac{M}{g} \tag{2}$$

The axial velocity of an extension wave can be expressed as [4],

$$V = L\sqrt{\frac{K}{m_0}} \tag{3}$$

where $K$ is the stiffness of the rainbow spring.

Based on Hooke's law [8], $K$ can be approximately computed as,

$$K = \frac{T}{L - L_0} \tag{4}$$

where $L_0$ is the length of the rainbow spring when the wires are compressed together. It should be noted that $L_0$ is the shortest length of the rainbow spring.

The time for the wave propagating the length of the spring can be expressed as,

$$t_w = \frac{L}{V} \tag{5}$$

Substituting Eqs. (1)-(4) into Eq. (5) gives,

$$t_w = \sqrt{\frac{\Delta L}{g}} \tag{6}$$

where $\Delta L = L - L_0$.

We will use this result to quantitatively predict the time before the overall fall of the rainbow spring.

Figure 2 shows the setup for the experiment. We use a high-speed camera 5F04M (Fuhuang Agile Device, China) to investigate the fall process of the rainbow spring. Two kinds of rainbow springs are considered, one plastic and another metal (see Figure 3A). For each rainbow spring, three situations are taken into account in terms of the loading condition, which are tabulated in Table 1. Static loads are produced by self-weight and/or adding extra weight using plastic tweezer. Two kinds of weight, namely 50 g and 100 g, are utilized (see Figure 3B). At the beginning, fix the spring carefully by hand with its axis vertically downward; the spring is statically stretched by self-weight and/or external weight. Then, release the spring and let it fall, and record its trajectory and the falling time until it hits the ground.

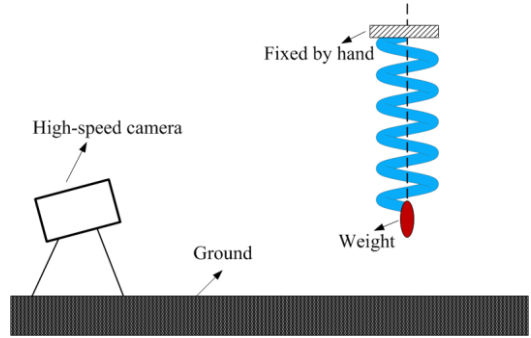

**FIG. 2.** Setup for the experiment.

**Table 1** Situations considered in the experiment

| Situations | Specific description |
| --- | --- |
| P[#1] | Plastic rainbow spring without external weight |
| P[#2] | Plastic rainbow spring with external weight ($M$=50 g) |
| P[#3] | Plastic rainbow spring with external weight ($M$=100 g) |
| M[#1] | Metal rainbow spring without external weight |
| M[#2] | Metal rainbow spring with external weight ($M$=50 g) |
| M[#3] | Metal rainbow spring with external weight ($M$=100 g) |

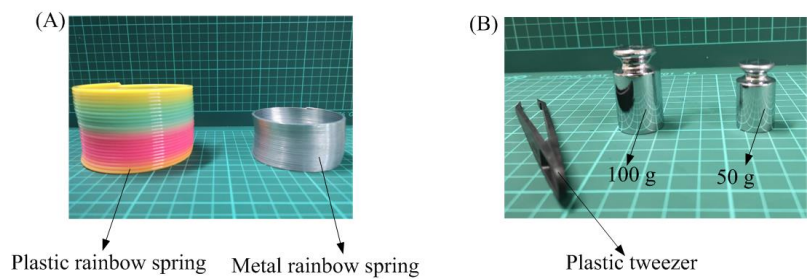

**FIG 3.** (A) Rainbow springs and (B) weigh utilized used in the experiment.

Figure 4 shows the experimental observations of the fall motions of the plastic rainbow spring with different $L$. The red dotted lines represent the resting position of the lower end of the plastic rainbow spring. $L_0$ equals to 0.056 m, and $L$ is 0.334 m, 0.778 m, and 1.225 m for situations P[#1], P[#2], and P[#3], respectively (see Supplementary Material, Movie S1, S2, S3). Situations P[#1], P[#2], and P[#3] are corresponding to Figure 4(A)-(C), respectively. Four close-ups extracted from the trajectory of the fall motion are depicted for each situation, including the initial static state, the intermediate state of the fall before the overall fall, the critical state before the overall fall, and the state of the overall fall, which correspond to the four photographs from left to right in Figure (A)-(C). From Figure 4, it can be clearly seen that, no matter what kinds of loading conditions, the lower end of the plastic rainbow spring holds still for a while as its upper parts move down, and then the whole spring falls together.

To further confirm the observation results, the fall motion of a metal rainbow spring is recorded and presented in Figure 5. The experimental conditions are identical to those of the plastic rainbow spring, except with different values of $L_0$ and $L$. $L_0$ equals to 0.037 m, and $L$ equals to 0.392 m, 0.759 m, and 1.108 m for situations $M^{\#1}$, $M^{\#2}$, and $M^{\#3}$ which are presented in Figure 5(A)-(C), respectively (see Supplementary Material, Movie S4, S5, S6). Experimental results show that the metal rainbow spring also shows the unconventional fall motion: before the overall metal rainbow spring falls together, its lower end the keeps still for some time with its upper parts moving down.

The motion of the upper part of the rainbow spring can be explained from the perspective of energy conversion. A certain potential energy is stored in the rainbow spring due to the extension when the spring is at rest in the initial stage. Potential energy is converted to kinetic energy after releasing the spring, leading to the motion of the upper part of the spring. However, it seems to be very difficult to give an explanation for the lower end of the spring staying still. We attempt to explain this unconventional phenomenon from the point of view of kinematics. We assume to cut the rainbow spring using a red dotted line, as shown in Fig. 1. The distance between the red dotted line and the top of the spring is $l$. The gravity of the wires above the red dotted line is $\bar{G} = \frac{l}{L}mg$ with $l < L$ and the static tension at the cutting end is assumed as $\bar{T}$. $\bar{T}$ varies gradually along the length of the spring. The reason for the motion of the upper part of the rainbow spring may be because $\bar{T}$ is always smaller than $\bar{G}$, making the wires of the upper part have a downward acceleration. In addition, before the upper part wires touch the bottom of the spring, the force at the bottom is balanced, therefore causing the lower end the keeps still for a while. However, when all the wires stack together (the spring has a length of $L_0$), which is a mutation process, the static tension disappears and the whole spring with the additional weight falls down together.

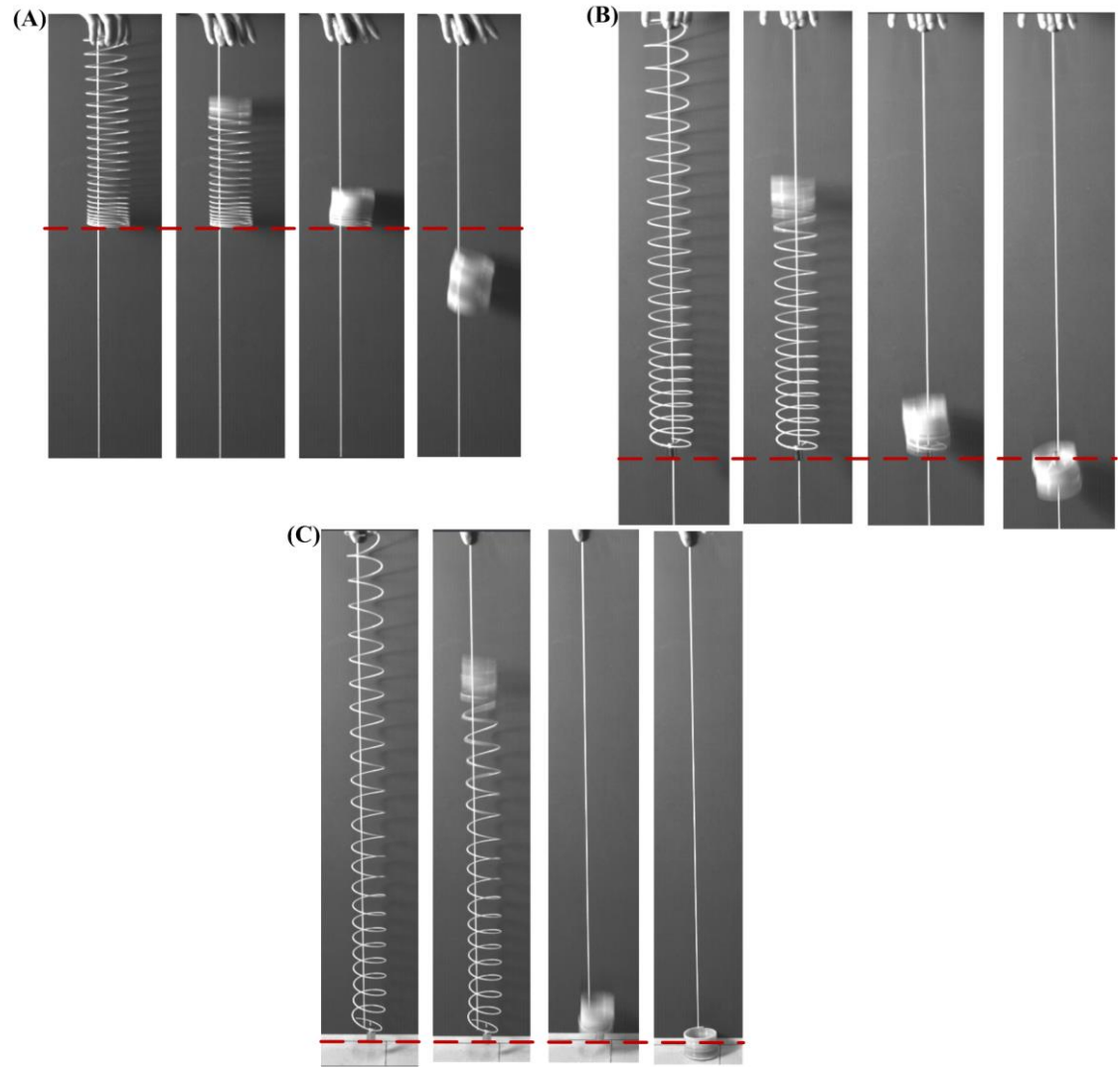

**FIG. 4.** Experimental observations of the fall motions of the plastic rainbow spring: (A) $P^{\#1}$, (B) $P^{\#2}$, and (C) $P^{\#3}$.

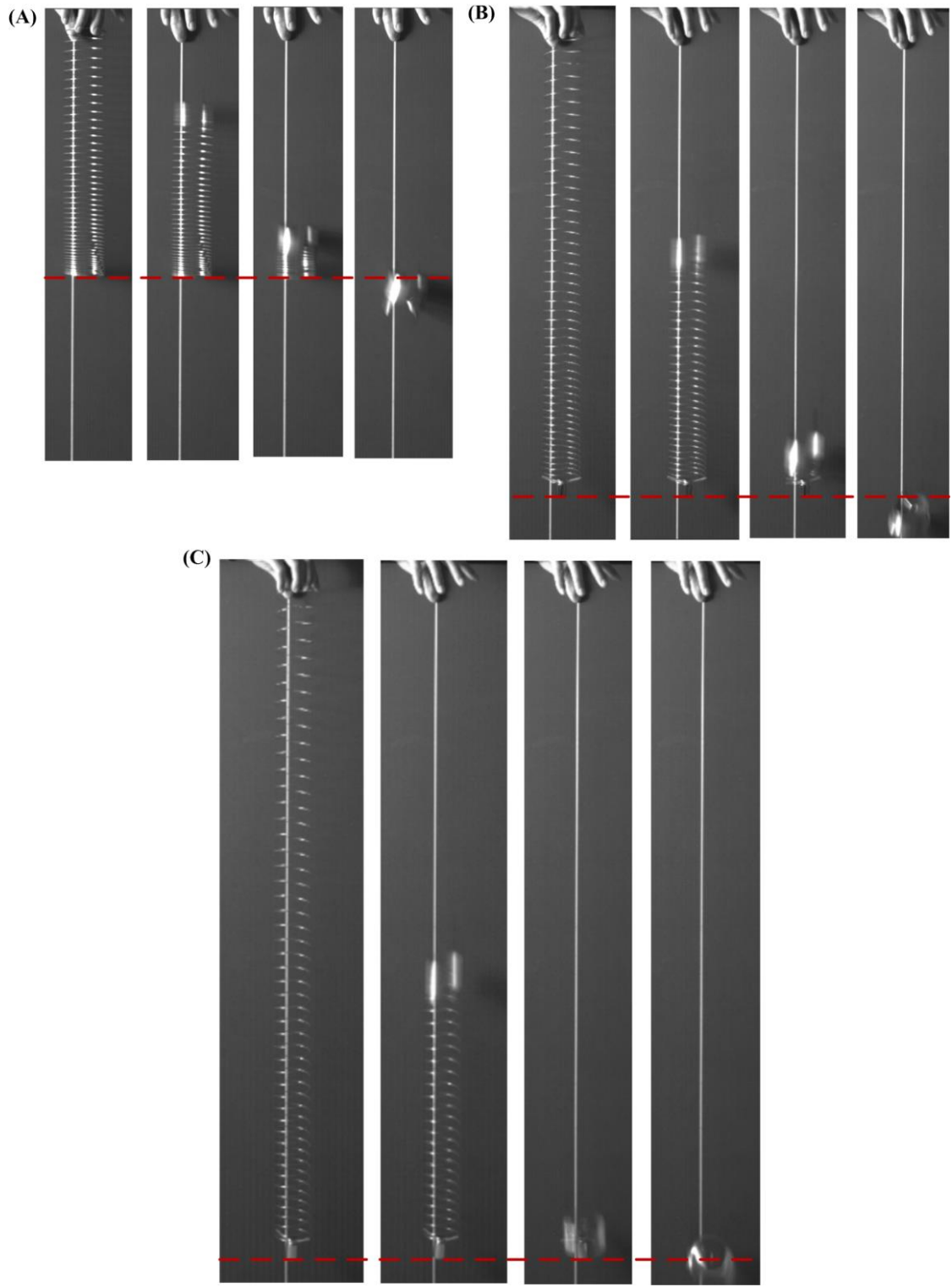

**FIG. 5.** Experimental observations of the fall motions of the metal rainbow spring: (A) M$^{\#1}$, (B) M$^{\#2}$, and (C) M$^{\#3}$.

Figure 6 shows that the time for the elastic wave propagating the length of the spring, $t_w$, is always larger than the time of the lower end holds still, $t$. Specifically, $t$ = 0.1667s, 0.1931s, and 0.2103s, respectively for P$^{\#1}$, P$^{\#2}$, and P$^{\#3}$ (highlighted by blue hollow square).

Correspondingly, $t_w$ = 0.1684s, 0.2714s, and 0.3454s. Similarly, for M$^{\#1}$, M$^{\#2}$, and M$^{\#3}$, $t$ = 0.1344s, 0.1689s, and 0.1724, marked by black hollow circle, and $t_w$ = 0.1903s, 0.2714s, and 0.3306s. Additional experimental tests with various ΔL are further conducted. Several values of $t$ extracted from the experiments are presented in Figure 6. It can be apparently seen that all $t$, no matter metal rainbow spring or the plastic rainbow spring, are below the curve of $t_w$. It means that $t$ is always smaller than $t_w$ for the same ΔL. This may be due to fact that, during the falling motion, the existed shock waves in the rainbow spring could accelerate its fall.

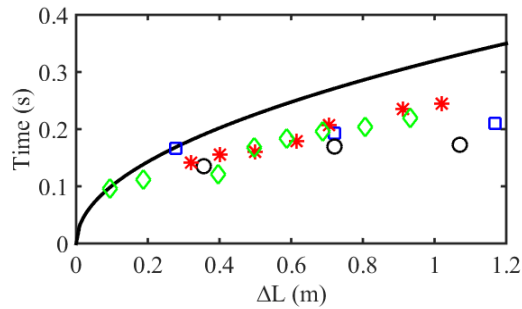

**FIG. 6.** Relationship between ΔL and $t_w$, $t$. Black line represents $t_w$, green hollow diamond represents plastic rainbow spring, and red asterisk stands for metal rainbow spring.

In summary, the unconventional fall motions of the metal and plastic rainbow springs are experimentally investigated using a high-speed camera. We have shown that, different from the traditional perception, after releasing from a high place, the lower end of the rainbow spring shows a temporary absolute space positioning such that the lower end first holds still for a while as its upper part moves down, and the whole spring does not fall down together before the wires of the upper parts stack together, which has been explained from the perspective of the energy conversion and kinematics. We have shown that the time of the lower end holds still, $t$, is always smaller the time for the elastic wave propagating the length of the spring, $t_w$ due to the existence of shock waves in the spring during its fall. The findings here can help us to better understand the physical falling motion properties of the objects and may have potential applications for the areas requiring temporary absolute space positioning, e.g., high resolution satellite positioning.

**Acknowledgements**:

This research was financially supported by the Key Program of National Natural Science Foundation of China (No. 11832009) and the Hong Kong Scholar Program (XJ2018052).